\documentstyle[epsfig,proceedings]{crckapb} 

\newcommand{\beq}{\begin{equation}}
\newcommand{\eeq}{\end{equation}}
\newcommand{\bd}{\begin{displaymath}}
\newcommand{\ed}{\end{displaymath}}
 
\newcommand{\p}{\partial}   
\newcommand{\Om}{\Omega}
\newcommand{\Omk}{\Omega_{\rm K}}
\newcommand{\rmd}{{\rm d}}

\newcommand{\aaa}[1]{{\it Astron.\ Astrophys.} {\bf #1}}

\newcommand{\apj}[1]{{\it Astrophys. J.} {\bf #1}}                    
          
\newcommand{\apjs}[1]{{\it Astrophys.\ J.\ Suppl.} {\bf #1}}

\newcommand{\mnras}[1]{{\it Mon.\ Not.\ R.\ astron.\ Soc.} {\bf #1}}

\newcommand{\pasj}[1]{{\it Publ.\ Astr.\ Soc.\ Japan} {\bf #1}}

\newcommand{\gapprox}{\;\rlap{\lower 2.5pt
             \hbox{$\sim$}}\raise 1.5pt\hbox{$>$}\;}
\newcommand{\lapprox}{\;\rlap{\lower 2.5pt
             \hbox{$\sim$}}\raise 1.5pt\hbox{$<$}\;}
\newcommand{\bfg}[1]{\setbox0=\hbox{#1}%
  \kern-.025em\copy0\kern-\wd0
  \kern.05em\copy0\kern-\wd0
  \kern-.025em\raise.0433em\box0}

\begin{opening}
  \title{Radiatively inefficient accretion disks}  
  \author{H.C. Spruit}
  \institute{Max-Planck-Institut f\"ur Astrophysik\\ Postfach 1523, D-85740 
Garching, Germany}
\end{opening}
\begin{document}
\begin{abstract}
Radiatively inefficient (or advection dominated) disks are 
discussed at an introductory level.  Ion supported and 
radiation supported flows are discussed, the different consequences of 
advection dominated flows onto black holes vs. solid surfaces (neutron stars, 
white dwarfs), hydrodynamics, the role of the ratio of specific heats, and 
the possible connection between ADAFs and outflows. 

To appear in `The neutron star black hole connection' (NATO ASI Elounda 1999,
eds. C. Kouveliotou and V. Connaughton).

\end{abstract}

\section{Introduction}
In a thin accretion disk, the time available for the accreting gas to radiate away 
the energy released by the viscous stress is the accretion time,
\beq t_{\rm acc}\approx{1\over\alpha\Omk}({r\over H})^2, \eeq
where $\alpha$ is the dimensionless viscosity parameter, $\Omk$ the local 
Keplerian rotation  rate, $r$ the distance from the central mass, and $H$ the 
disk thickness (see Frank et al. 1992 or Spruit, elsewhere in this volume). For a 
thin disk, $H/r\ll 1$, this time is much longer than the thermal time scale 
$t_{\rm t}\approx 1/(\alpha\Om)$. There is then enough time for a local balance to 
exist between viscous dissipation and radiative cooling. For the accretion rates 
implied in observed systems the disk is then rather cool, and the starting 
assumption $H/r\ll 1$ is justified. 

This argument is somewhat circular, however, since the accretion time is long 
enough for effective cooling only if the disk is assumed to be thin to begin with. 
Other forms of accretion disks may exist, even at the same accretion rates, in 
which the cooling is ineffective compared with that of standard (geometrically 
thin, optically thick) disks. 

Since radiatively inefficient disks tend to be thick, $H/r\sim O(1)$, they are 
sometimes called `quasi-spherical'.  However, this does {\em not} mean that a 
spherically symmetric accretion model would be a reasonable approximation. 
The crucial difference is that the flow has angular momentum. The inward flow 
speed is governed by the rate at which angular momentum can be transferred 
outwards, rather than by gravity and pressure gradient. The accretion time 
scale, $t_{\rm acc}\sim 1/(\alpha\Om)$ is longer than the accretion time scale 
in the spherical case (unless the viscosity parameter $\alpha$ is as large as 
$O(1)$). The dominant velocity component is azimuthal rather than radial, and 
the density and optical depth are much larger than in the spherical case.

It turns out that there are two kinds of radiatively inefficient disks, the 
optically thin, and optically thick varieties. A second distinction occurs because 
accretion flows are different for central objects with a solid surface (neutron 
stars, white dwarfs, main sequence stars, planets), and those without (i.e. black 
holes). I start with optically  thick flows. 

\section{Radiation supported advective accretion}
\label{radv}
If the energy loss by radiation is small, the gravitational energy release 
$W_{\rm grav}\approx GM/(2r)$ is converted into enthalpy of the gas and 
radiation field\footnote{I assume here that a fraction $\sim 0.5$ of the gravitational potential energy stays in the flow as orbital kinetic energy. See also section 3.}
\beq 
{1\over 2}{GM\over r}={1\over\rho}[{\gamma\over\gamma-1}P_{\rm g}+ 
4P_{\rm r}], \label{tv}
\eeq
where an ideal gas of constant ratio of specific heats $\gamma$ has been 
assumed, and $P_{\rm r}={1\over 3}aT^4$ is the radiation pressure. In terms of 
the virial temperature $T_{\rm vir}=GM/({\cal R}r)$, and assuming $\gamma=5/3$, 
appropriate for a fully ionized gas, this can be written as 
\beq {T\over T_{\rm vir}}=[5+8{P_{\rm r}\over P_{\rm g}}]^{-1}. \label {ttv} \eeq
Thus, for radiation pressure dominated accretion, $P_{\rm r}\gg P_{\rm g}$ , the 
temperature is much less than the virial temperature. The disk thickness is 
given by
\beq  H\approx [(P_{\rm g}+P_{\rm r})/\rho]^{1/2}/\Om,\eeq
With (\ref{ttv}) this yields
\beq H/r\sim O(1).\eeq
In the limit $P_{\rm r}\gg P_{\rm g}$, the flow is therefore geometrically thick. 
This implies that radiation pressure supplies a non-negligible fraction of the 
support of the gas against gravity (the remainder being provided by rotation). 

For $P_{\rm r}\gg P_{\rm g}$, (\ref{tv}) yields
\beq {GM\over 2r}={4\over 3}{aT^4\over\rho}.\label{vir}\eeq
The radiative energy flux, in the diffusion approximation, is
\beq F={4\over 3}{\rmd\over \rmd\tau}\sigma T^4\approx {4\over 3}{\sigma 
T^4\over \tau},\eeq
where $\sigma=ac/4$ is Stefan-Boltzmann's radiation constant. Hence
\beq  F={1\over 8}{GM\over rH}{c\over\kappa}=F_{\rm E}{r\over 8H},\label{flux}\eeq
where $F_{\rm E}=L_{\rm E}/(4\pi r^2)$ is the local Eddington flux. Since 
$H/r\approx 1$, a radiatively inefficient, radiation pressure dominated accretion 
flow has a luminosity of the order of the Eddington luminosity.

The temperature depends on the accretion rate and the viscosity $\nu$ assumed. 
The accretion rate is of the order $\dot M\sim 3\pi\nu\Sigma$ (see `accretion 
disks' elsewhere in this volume), where $\Sigma=\int\rho\rmd z$ is the surface  
mass density. In units of the Eddington rate, we get
\beq \dot m\equiv\dot M/\dot M_{\rm E}\approx {\nu\rho\kappa/c}, \eeq
where $H/r\approx 1$ has been used, and $\dot M_{\rm E}$ is the Eddington 
accretion rate onto the central object of size $R$,
\beq 
\dot M_{\rm E}={R\over GM}L_{\rm E}= 4\pi Rc/\kappa. \label{medd}
\eeq
[Note that the definition of $\dot M_{\rm E}$ differs by factors of order unity 
between different authors. It depends on the assumed efficiency $\eta$ of 
conversion of gravitational energy $GM/R$ into radiation. In (\ref{medd}) it is 
taken to be unity, for accretion onto black holes a more realistic value is $\eta=0.1$, for 
accretion onto neutron stars $\eta\approx 0.4$.] 

Assume that the viscosity scales with the  gas pressure:
\beq \nu=\alpha{P_{\rm g}\over\rho\Omk}, \eeq
instead of the total pressure $P_{\rm r}+P_{\rm g}$. This is the form that is 
likely to hold if the angular momentum transport is due to a small-scale 
magnetic field (Sakimoto and Coroniti, 1989). Then with (\ref{vir}) we have (up to a factor $2H/r\sim O(1)$
\beq 
T^5\approx {(GM)^{3/2}\over r^{5/2}}{{\dot m c}\over\alpha\kappa a{\cal R}}, 
\eeq
or
\beq T\approx 2\,10^8 r_6^{-1/5}( r/r_{\rm g})^{3/10}\dot m^{1/5}, \eeq
where $r=10^6r_6$ and $r_{\rm g}=2GM/c^2$ is the gravitational radius of the 
accreting object, and the electron scattering opacity of 0.3 cm$^2$/g has been assumed. The temperatures expected in radiation supported advection 
dominated flows are therefore quite low compared with the virial temperature 
[If the viscosity is assumed to scale with the total pressure instead of $P_{\rm 
g}$, the temperature is even lower]. The effect of electron-positron pairs can be 
neglected (Schultz and Price, 1985), since they are present only at 
temperatures approaching the electron rest mass energy, $T\gapprox 10^9$K.

In order for the flow to be radiation pressure and advection dominated, the 
optical depth has to be sufficiently large so the radiation does not leak out. The 
energy density in the flow, vertically integrated at a distance $r$, is of the 
order
\beq E\approx aT^4H,\eeq
and the energy loss rate per cm$^2$ of disk surface is given by (\ref{flux}).
The cooling time is therefore,
\beq t_{\rm c}=E/F=3\tau H/c.\eeq
This is to be compared with the accretion time, which can be written in terms 
of the mass in the disk at radius $r$, of the order $2\pi r^2\Sigma$, and the 
accretion rate:
\beq t_{\rm acc}=2\pi r^2\Sigma/\dot M.\eeq
This yields
\beq 
t_{\rm c}/t_{\rm acc}\approx{\kappa\over\pi rc}\dot M=4\dot m{R\over r}, 
\eeq
(where a factor $3/2\,H/r\sim O(1)$ has been neglected). Since $r>R$, this shows 
that accretion has to be super-Eddington in order to be both radiation- and 
advection-dominated. 

This condition can also be expressed in terms of the so-called {\em trapping 
radius} $r_{\rm t}$ (e.g. Rees 1978). Equating $t_{\rm acc}$ and $t_{\rm c}$ yields
\beq r_{\rm t}/ R\approx 4 \dot m. \eeq
Inside $r_{\rm t}$, the flow is advection dominated: the radiation field produced 
by viscous dissipation stays trapped inside the flow, instead of being radiated 
from the disk as happens in a standard thin disk. Outside the trapping radius, the 
radiation field can not be sufficiently strong to maintain a disk with $H/r\sim 
1$, it must be a thin form of disk instead. Such a thin disk can still be 
radiation-supported (i.e. $P_{\rm r}\gg P_{\rm g}$), but it can not be advection 
dominated.

Flows of this kind are called `radiation supported tori' (or radiation tori, for 
short) by Rees et al. 1982\footnote{Some workers interpret the use of word 
`torus' in the context radiatively inefficient accretion as implying a rotating 
but non-accreting flow. The physics studied by Rees et al. (1982), however, where 
this name was introduced, explicitly refers to accreting flows such as 
described here}. They must accrete at a rate above the Eddington value to exist. 
The converse is not quite true: a flow accreting above Eddington is an advection 
dominated flow, but it need not necessarily be radiation dominated.  Advection 
dominated optically thick acretion flows exist in which radiation does not play 
a major role (see section \ref{planet}). 

That an accretion flow above $\dot 
M_{\rm E}$ is advection dominated, not a thin disk, follows from the fact that 
in a thin disk the energy dissipated must be radiated away locally. Since the 
local radiative flux can not exceed the Eddington energy flux $F_{\rm E}$, the 
mass accretion rate in a thin disk can not significantly exceed the Eddington value 
(\ref{medd}).

The gravitational energy, dissipated by viscous stress in differential rotation 
and advected with the flow, ends up at the central object. If this is a black hole, 
the photons, particles and their thermal energy are conveniently swallowed at 
the horizon, and do not react back on the flow. Radiation tori are therefore 
mostly relevant for accretion onto black holes. They are convectively unstable 
(Bisnovatyi-Kogan and Blinnikov 1977): the way in which energy is dissipated, 
in the standard $\alpha$-prescription, is such that the entropy ($\sim 
T^3/\rho$) decreases with height in the disk. Recent numerical simulations 
(see section \ref{outflows}) show the effects of this convection.

\subsection{Super-Eddington accretion onto black holes}
As the accretion rate onto a black hole is increased above $\dot M_{\rm E}$, the 
trapping radius moves out. The total luminosity increases only slowly, and 
remains of the order of the Eddington luminosity. Such supercritical accretion 
has been considered by Begelman and Meier (1982, see also Wang and Zhou 
1999); they show that the flow has a radially self-similar structure.

Abramowicz et al. (1988, 1989) studied accretion onto black holes at rates near 
$\dot M_{\rm E}$. They used a vertically-integrated approximation for the disk, 
but included the advection terms. The resulting solutions were called `slim 
disks'. These models show how with increasing accretion rate, a standard thin 
Shakura-Sunyaev disk turns into a radiation-supported advection flow. The 
nature of the transition depends on the viscosity prescription used, and can 
show a non-monotonic dependence of $\dot M$ on surface density $\Sigma$ 
(Honma et al. 1991). This suggests the possibility of instability and cyclic behavior 
of the inner disk near a black hole,  at accretion rates near and above $\dot 
M_{\rm E}$ (for an application to GRS 1915+105 see Nayakshin et al., 1999). 

\subsection{Super-Eddington accretion onto neutron stars}
In the case of accretion onto a neutron star, the energy trapped in the flow, plus the 
remaining orbital energy, settles onto its surface. If the accretion rate is below 
$\dot M_{\rm E}$, the energy can be radiated away by the surface, and steady 
accretion is possible. A secondary star providing the mass may, under some 
circumstances, transfer more than $\dot M_{\rm E}$, since it does not 
know about the neutron star's Eddington value. The outcome of this case 
is still somewhat uncertain; it is generally believed on intuitive grounds that 
the `surplus' (the amount above $\dot M_{\rm E}$) somehow gets expelled from 
the system. 

As the transfer rate is increased, the accreting hot gas forms an 
extended atmosphere around the neutron star, like the envelope of a giant.  If it 
is large enough, the outer parts of this envelope are partially ionized. The 
opacity in these layers, due to lines of the CNO and heavier elements, is then 
much higher than the electron scattering opacity. The Eddington luminosity 
based on the local value of the opacity is then smaller than it is near the neutron 
star surface. Once an extended atmosphere with a cool surface forms, the 
accretion luminosity is thus large  enough to drive a wind from the envelope 
(see Kato 1997, where this is discussed in the context of Novae). 

This scenario 
is somewhat dubious however, since it assumes that the mass transferred from 
the secondary continues to reach the neutron star and generate a high 
luminosity there. This is not at all obvious, since the mass transfering stream 
may dissipate in the growing neutron star envelope instead. The result would be 
a giant  (more precisely, a Thorne-Zytkow star), with a steadily increasing 
envelope mass. Such an envelope is likely to be large enough to envelop the 
entire binary system, which then develops into a common-envelope (CE) system. The 
envelope mass is expected to be ejected by CE hydrodynamics (Taam 1994, 2000). 

A more 
speculative proposal, suggested by the properties of SS 433, is that the `surplus 
mass' is ejected in the form of jets. 
The binary parameters of Cyg X-2 are observational evidence for mass ejection 
in super-Eddington mass transfer phases (King and Ritter 1999, Rappaport and 
Podsiadlowski 1999, King and Begelman 1999). 

\section{Hydrodynamics}
The hydrodynamics of ADAFs and radiation tori can be studied by starting, at a 
very simple level, with a generalization of the thin disk equations. Making the 
assumption that quantities integrated over the height $z$ of the disk give a fair 
representation (though this is justifiable only for thin disks), and assuming 
axisymmetry, the problem reduces to a one-dimensional time-dependent one. 
Further simplifying this by restriction to a steady flow yields the equations
\beq 2\pi r\Sigma v_r=\dot M={\rm cst} \label{cnt},\eeq
\beq 
r\Sigma v_r\p_r(\Om r^2)=\p_r(\nu\Sigma r^3\p_r\Om),\label{ang}
\eeq
\beq 
v_r\p_r v_r-(\Om^2-\Om^2_{\rm K})r=-{1\over\rho}\p_r p,\label{vr}
\eeq 
\beq\Sigma v_rT\p_rS=q^+-q^-, \label{eng}\eeq
where $S$ is the specific entropy of the gas,  $\Om$ the local rotation rate, 
now different from the Keplerian rate $\Omk=(GM/r^3)^{1/2}$, while 
\beq 
q^+=\int Q_{\rm v}\rmd z  \qquad q^-=\int {\rm div}F_{\rm r}\rmd z \label{qs}
\eeq
are the height-integrated viscous dissipation rate and radiative loss rate, 
respectively. In the case of thin disks, equations (\ref{cnt}) and (\ref{ang}) are 
unchanged, but (\ref{vr}) simplifies to $\Om^2=\Omk^2$, i.e. the rotation is 
Keplerian, while (\ref{eng}) simplifies to $q^+=q^-$, expressing local balance 
between viscous dissipation and cooling. The left hand side of (\ref{eng}) 
describes the radial advection of heat, and is perhaps the most important 
deviation from the thin disk equations at this level of approximation (hence the 
name advection dominated flows). The characteristic properties are seen most 
cearly when radiative loss is neglected altogether, $q^-=0$. 
The equations are supplemented with expressions for $\nu$ and $q^+$:
\beq 
\nu=\alpha c_{\rm s}^2/\Omk; \qquad q^+=(r\p_r\Om)^2\nu\Sigma.
\eeq
If $\alpha$ is taken constant, $q^-=0$, and an ideal gas is assumed with 
constant ratio of specific heats, so that the entropy is given by
\beq S=c_{\rm v}\ln (p/\rho^\gamma), \eeq
then equations (\ref{cnt})-(\ref{eng}) have no explicit length scale in them. This 
means that a special so-called self-similar solution exists, in which all 
quantities are powers of $r$.  Such self-similar solutions have apparently first 
been described by Gilham (1981), but have since then been re-invented several 
times (Spruit et al. 1987; Narayan and Yi, 1994). The dependences on $r$  
are 
\beq \Om\sim r^{-3/2};\qquad \rho\sim r^{-3/2},\eeq
\beq H\sim r; \qquad T\sim r^{-1}.\eeq
In the limit $\alpha\ll 1$, one finds
\beq 
v_r=-\alpha\Omk r \left(9{\gamma-1\over 5-\gamma}\right),
\eeq
\beq 
\Om=\Omk\left(2{5-3\gamma\over 5-\gamma}\right)^{1/2},
\eeq
\beq c_{\rm s}^2=\Omk^2r^2{\gamma-1\over 5-\gamma}, \eeq
\beq {H\over r}=\left(\gamma-1\over 5-\gamma\right)^{1/2}.\eeq
The precise from of these expressions depends somewhat on the way in which vertical integrations such as in (\ref{qs}) are done (which are only approximate).

The self-similar solution can be compared with numerical solutions of eqs. 
(\ref{cnt})--(\ref{eng}) with appropriate conditions applied at inner ($r_{\rm i}$) 
and outer ($r_{\rm o}$) boundaries (Nakamura et al. 1996, Narayan et al. 1997). 
The results show that the self-similar solution is valid in an intermediate 
regime $r_{\rm i}\ll r\ll r_{\rm o}$. That is, the solutions of 
(\ref{cnt})--(\ref{eng}) approach the self-similar solution far from the 
boundaries, as is characteristic of self-similar solutions.

The solution exists only if $1<\gamma\le 5/3$, a condition satisfied by all 
ideal gases. As $\gamma\downarrow 1$, the disk temperature and thickness 
vanish. This is understandable, since a $\gamma$ close to 1 means that the gas 
has a large number of internal degrees of freedom. The accretion energy is 
shared between all degrees of freedom, so that for a low $\gamma$ less is 
available for the kinetic energy (temperature) of the particles. 

Second, the {\em rotation rate vanishes} for $\gamma\rightarrow 5/3$. Since a fully 
ionized gas has $\gamma=5/3$, it is the most relevant value for optically thin 
accretion near a black hole or neutron star. Apparently, steady advection 
dominated accretion can not have angular momentum in this case, and the
question arises how an adiabatic flow with angular momentum will behave for  $\gamma=5/3$. In the literature, this problem has been circumvented by arguing 
that real flows would have magnetic fields in them,  which would change the effective 
compressibility of the gas. Even if a magnetic field of sufficient strength is 
present, however, (energy density comparable to the gas pressure) the effective 
$\gamma$ is not automatically lowered. If the field is compressed mainly 
perpendicular to the field lines, for example, the effective $\gamma$ 
is closer to 2. Also, this does not solve the conceptual problem what would happen to a 
rotating accretion flow consisting of a more weakly magnetized ionized gas. 

This conceptual problem has been solved by Ogilvie (1999), who showed how the 
low rotation for $\gamma=5/3$ comes about in a time-dependent manner. He 
found a similarity solution (depending on distance and time 
in the combination $r/t^{2/3}$) to the time-dependent version of eqs 
(\ref{cnt})--(\ref{eng}). This solution describes the asymptotic behavior (in 
time) of a viscously spreading disk, analogous to the viscous spreading of thin 
disks (Lynden-Bell and Pringle 1972). As in the thin disk case, all the mass 
accretes asymptotically onto the central mass, while all the angular momentum 
travels to infinity together with a vanishing amount of mass. For all 
$\gamma<5/3$, the rotation rate at a fixed $r$ tends to a finite value as 
$t\rightarrow\infty$, but for $\gamma=5/3$ it tends to zero. The 
slowly-rotating region expands in size as $r\sim t^{2/3}$. It thus seems likely 
that the typical slow rotation of ADAFs at $\gamma$ near 5/3 is a real physical 
property, and that the angular momentum gets expelled from the inner regions of the flow.

\subsection{Other optically thick accretion flows}
\label{planet}
The radiation-dominated flows discussed in section \ref{radv} are not the only 
possible optically thick advection dominated flows. From the discussion of the 
hydrodynamics, it is clear that disk-like (i.e. rotating) accretion is possible 
whenever the ratio of specific heats is less than 5/3. A radiation supported 
flow satisfies this requirement with $\gamma=4/3$, but it can also happen in 
the absence of radiation if energy is taken up in the gas by internal degrees of 
freedom of the particles. Examples are the rotational and vibrational degrees of 
freedom in molecules, and the energy associated with dissociation and ionization. 
If the accreting object has a gravitational potential not much exceeding the 13.6 + 
2.2 eV per proton for dissociation plus ionization, a gas initially consisting of 
molecular  hydrogen 
can stay bound at arbitrary accretion rates. This translates into a limit 
$M/M_\odot\, R_\odot/R<0.01$. This is satisfied approximately by the giant 
planets, which are believed to have gone through a phase of rapid adiabatic gas 
accretion  (e.g. Podolak et al. 1993). 

A more remotely related example is the core-collapse supernova. The accretion 
energy of the envelope mass falling onto the proto-neutron star is lost mostly 
through photodisintegration of nuclei, causing the well known problem of 
explaining how a shock is produced of sufficient energy to unbind the envelope. 
If the pre-collapse core rotates sufficiently rapidly, the collapse will form an 
accretion  torus (inside the supernova envelope), with properties similar to 
advection dominated accretion flows (but at extreme densities and accretion 
rates, by X-ray binary standards). Such objects have been invoked as sources of 
Gamma-ray bursts (Popham et al. 1999, see also the review by Meszaros, 
elsewhere in this volume).

A final possibility for optically thick accretion is through {\em neutrino 
losses}. If the temperature and density near an accreting neutron star become  
large enough, additional cooling takes place through neutrinos (as in the cores 
of giants). This is relevant for the physics of Thorne-Zytkow stars (neutron 
stars or black holes in massive supergiant envelopes, cf.\ Bisnovatyi-Kogan and 
Lamzin 1984, Cannon et al. 1992), and perhaps for the spiral-in of neutron stars 
into giants (Chevalier 1993, see however Taam 2000).

\section{Optically thin advection dominated flows (ADAFs)}
The optically thin case has received most attention in recent years, because of 
the promise it holds for explaining the (radio to X-ray) spectra of X-ray binaries and the 
central black holes in galaxies, including our own.  For a recent review see Yi (1999). 
This kind of flow occurs if the gas is optically thin, and radiation processes 
sufficiently weak. The gas then heats up to near the virial temperature. Near the 
last stable orbit of a black hole, this is of the order 100 MeV, or $10^{12}$K.  At 
such temperatures, a gas in thermal equilibrium would radiate at a fantastic 
rate, even if it were optically thin, because the interaction between electrons 
and photons becomes very strong already near the electron rest mass of 0.5MeV. 
In a remarkable paper, Shapiro Lightman and Eardley (1976) noted that this, 
however, is not what will happen in an optically thin accreting plasma but that, 
instead, a {\it two-temperature plasma} forms. 

Suppose that the energy released by viscous dissipation is distributed equally 
among the carriers of mass, i.e. mostly to the ions and $\sim 1/2000$ to the 
electrons. Most of the energy resides in the ions, which radiate very 
inefficiently (their high mass prevents the rapid accelerations that are needed 
to produce electromagnetic radiation). This energy is transfered to the 
electrons by Coulomb interactions. These interactions are slow, however, 
under the conditions mentioned. They are slow because of the low density (on 
account of the assumed optical tickness), and because they decrease with increasing 
temperature. The electric forces that transfer energy from an ion to an electron act 
only as long as the ion is within the electron's Debye sphere (e.g. Spitzer, 1965). The 
interaction time between proton and electron, and thus the momentum transfered, 
therefore decrease as $1/v_{\rm p}\sim T_{\rm p}^{-1/2}$ where $T_{\rm p}$ is 
the proton temperature. 

In this way, an optically thin plasma near a compact object can be in a 
two-temperature state, with the ions being near the virial temperature, and 
the electrons, which are doing the radiating, at a much lower temperature 
around 50--200 keV. The energy transfer from the gravitational field to the 
ions is fast (by some form of viscous or magnetic dissipation), from the ions to 
the electrons slow, and finally the energy losses of the electrons fast (by 
synchrotron radiation in a magnetic field or by inverse Compton scattering of soft 
photons). Such a flow would be radiatively inefficient since the receivers of the 
accretion energy, the ions, get swallowed by the hole before getting a chance to 
transfer their energy to the electrons. The first disk models which take into 
account the physics of advection and a two-temperature plasma were developed by Ichimaru  (1977).

It is clear from this description that both the physics of such flows and the 
radiation spectrum to be expected depend crucially on the 
details of the ion-electron interaction and radiation processes assumed. This is 
unlike the case of the optically thick advection dominated flows, where gas 
and radiation are in approximate thermodynamic equilibrium. This is a source of 
uncertainty in the application of the optically thin ADAFs to observed systems, 
since their radiative properties depend on poorly known quantities, for example, the 
strength of the magnetic field in the flow.

The various branches of optically thin and thick accretion flows are summarized 
in figure 1. Each defines a relation between surface density $\Sigma$ (or 
optical depth $\tau=\kappa\Sigma$) and accretion rate. Optically thin ADAFs 
require low densities, either  because of low accretion rates or large values of 
the viscosity parameter. The condition that the cooling time of the ions by 
energy transfer to the electrons is longer than the accretion time yields a 
maximum accretion rate (Rees et al. 1982), 
\beq \dot m\lapprox \alpha^2. \eeq
If $\alpha\approx 0.05$ as suggested by current simulations of magnetic turbulence, the 
maximum accretion rate would be a few $10^{-3}$. If ADAFs are to be applicable 
to systems with higher accretion rates, such as Cyg X-1 for example, the 
viscosity parameter must be larger, on the order of 0.3. 

\begin{figure}
\mbox{}\hfill\epsfysize8cm\epsfbox{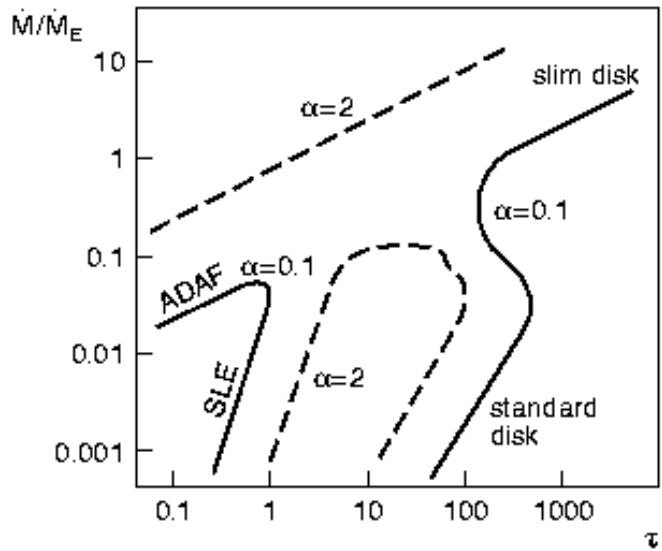}\hfill\mbox{}
\caption{Branches of advection-dominated and thin disks for two values or the 
viscosity parameter $\alpha$, as functions of accretion rate and (vertical) 
optical depth of the flow (schematic, after Chen et al. 1995, Zdziarski 1998). 
Optically thin branches are the ADAF and SLE (Shapiro-Lightman-Eardley) 
solutions, optically thick ones the radiation dominated (`slim disk' or `radiation 
torus') and SS (Shakura-Sunyaev or standard thin disk). Advection dominated are 
the ADAF and radiation torus,  geometrically thin are the SLE and SS. The SLE 
solution is a thermally unstable branch.}
\end{figure}

\subsection{Application: hard spectra in X-ray binaries}
In the hard state, the X-ray spectrum of black hole and 
neutron star accreters is characterized by a peak in the energy distribution 
($\nu F_\nu$ or $E\,F(E)$) at photon energies around 100 keV. This is to be 
compared with the typical photon energy of $\sim 1$ keV expected from a 
standard optically thick thin disk accreting near the Eddington limit. The 
standard, and by far most likely explanation is that the observed hard photons 
are softer photons (around 1 keV) that have been up-scattered through inverse 
Compton scattering on hot electrons. Fits of such Comptonized spectra (e.g. 
Zdziarski 1998 and references therein) yield an electron scattering optical 
depth around unity and an electron temperature of 50--100 keV. The scatter in 
these parameters is rather small between different sources. The reason may lie 
in part in the physics of Comptonization, but is not fully understood either. 
Something in the physics of the accretion flow keeps the Comptonization 
parameters constant as long as it is in the hard state. ADAFs have been applied 
with some success in interpreting XRB. They can produce reasonable X-ray 
spectra, and have been used in interpretations of the spectral-state 
transitions in sources like Cyg X-1 (Esin et al. 1998 and references therein).

An alternative to the ADAF model for the hard state in sources like Cyg X-1 and 
the black hole X-ray transients is the `corona' model. A hot corona 
(Bisnovatyi-Kogan and Blinnikov 1976), heated perhaps by magnetic fields as in 
the case of the Sun  (Galeev et al. 1979) could be the medium that Comptonizes 
soft photons radiated from the cool disk underneath. The energy balance in such 
a model produces a Comptonized spectrum within the observed 
range (Haardt and Maraschi 1993). This model has received further momentum, 
especially as a model for AGN, with the discovery of broadened X-ray lines 
indicative of the presence of a cool disk close to the last stable orbit around a 
black hole (Lee et al. 1999 and references therein). The very rapid X-ray 
variability seen in some of these sources is interpreted as due magnetic flaring 
in the corona, like in the solar corona (e.g. Di Matteo et al. 1999a).

\subsection{Transition from thin disk to ADAF}
One of the difficulties in applying ADAFs to specific observed systems is the 
transition from a standard geometrically thin, optically thick disk, which must 
be the mode of mass transfer at large distances, to an ADAF at closer range. 
This is shown by figure 1, which illustrates the situation at some distance 
close to the central object. The standard disk and the optically thin branches 
are separated from each other for all values of the viscosity parameter. This 
separation of the optically thin solutions also holds at larger distances. Thus, 
there is no plausible continuous path from one to the other, and the transition 
between the two must be due to additional physics that is not included in 
diagrams like figure 1. 

A promising possibility is that the transition takes 
place through a gradual (as a function of radius) {\em evaporation} (Meyer and 
Meyer-Hofmeister 1994, Meyer-Hofmeister and Meyer 1999). In this scenario,  
evaporation initially produces a corona above the disk, which transforms into an 
ADAF further in.

\subsection{Quiescent galactic nuclei}
For very low accretion rates, such as inferred for the black hole in the center of 
our galaxy (identified with the radio source Sgr A*), the broad band spectral 
energy distribution of an ADAF is 
predicted to have two humps (Narayan et al. 1995, Quataert et al. 1999). In the 
X-ray range, the emission is due to bremsstrahlung. In the radio range, the flow 
emits synchrotron radiation, provided that the magnetic field in the flow has an 
energy density order of the gas pressure (`equipartition'). Synthetic ADAF spectra 
can be fitted to 
the observed radio and X-ray emission from Sgr A*. In other galaxies where a 
massive central black hole is inferred, and the center is populated by an X-ray 
emitting gas of known density, ADAFs would also be natural, and might 
explain why the observed luminosities are so low compared with the 
accretion rate expected for a hole embedded in a  gas of the measured density. 
In some of these galaxies, however, the peak in the radio-to-mm range 
predicted by analogy with Sgr A* is not observed (Di Matteo et al. 1999b). This 
requires an additional hypothesis, for example that the magnetic field in these 
cases is much lower, or that the accretion energy is carried away by an outflow. 

\subsection{Transients in quiescence}
X-ray transients in quiescence (i.e. after an outburst) usually show a very low 
X-ray luminosity. The mass transfer rate from the secondary in quiescence can 
be inferred from  the optical emission. This shows the characteristic `hot spot', 
known from other systems to be the location where the mass transfering 
stream impacts on the edge of an accretion disk (e.g. van Paradijs and 
McClintock 1995). These observations thus show that a disk is present in 
quiescence, while the mass transfer rate can be measured from the brightness 
of the hot spot. If this disk were to extend to the neutron star with constant mass 
flux, the predicted X-ray luminosity would be much higher than observed. This 
has traditionally been interpreted as a consequence of the fact that in transient 
systems, the accretion is not steady. Mass is stored in the outer 
parts and released  by a disk instability (e.g. King 1995, Meyer-Hofmeister and 
Meyer 1999) producing the X-ray outburst. During quiescence, the accretion
rate onto the compact object is much smaller than the mass transfer from the 
secondary to the disk.

ADAFs have been invoked as an alternative explanation. The quiescent accretion 
rate onto the central object is proposed to be higher than in the disk-instability 
explanation, the greater energy release being hidden on account of the low 
radiative efficiency of the ADAF. Some transient systems have neutron star 
primaries, with a hard surface at 
which the energy  accreted by the ADAF must somehow be radiated away. A 
neutron star, with or without ADAFs, can not accrete in a 
radiatively inefficient way. In order to make ADAFs applicable, it has been 
proposed that the neutron stars in these systems have a modest magnetic dipole 
moment, such that in quiescence the gas in the accretion disk is prevented, by 
the `propeller effect' (Illarionov and Sunyaev 1975, Sunyaev and Shakura 1977) 
from accreting onto the star.

\subsection{ADAF-disk interaction: Lithium}
One of the strong predictions of ADAF models, whether for black holes or 
neutron stars, is that the accreting plasma in the inner regions has an ion 
temperature of 10--100 MeV. Nearby is a cool and dense accretion disk feeding 
this plasma. If only a small fraction of the hot ion plasma gets in contact with 
the disk, the intense irradiation by ions will produce nuclear reactions 
(Aharonian and Sunyaev 1984, Mart\'{\i}n et al. 1992). The main effects would 
be spallation of CNO elements into Li and Be, and the release of neutrons by 
various reactions. In this context, it is intriguing that the secondaries of both 
neutron star and black hole accreters have high overabundances of Li compared 
with other stars of their spectral types (Mart\'{\i}n et al. 1992, 1994a). If a 
fraction of the disk material is carried to the secondary by a disk wind, the 
observed Li abundances may be accounted for (Mart\'{\i}n et al. 1994b). 

\subsection{ADAF-disk interaction: hard X-spectra}
The interaction of a hot ion plasma with the cool disk produces a surface layer 
heated by the incident ions through Coulomb interactions with the electron gas 
in the disk. Its thickness and temperature turn out to be largely self-regulating: 
the energy balance is as in the Haardt and Maraschi corona models (in which the 
interaction is by photons only), while the optical thickness self-regulates 
through the dependence of the ion penetration depth on the electron 
temperature. This model (Spruit 1997) produces hard comptonized X-ray spectra 
whose shape is largely independent of both the energy flux and distance from 
the central object.

\section{Outflows?}
\label{outflows}
The energy density in an advection dominated accretion  flow is of the same 
order as the gravitational binding energy density $GM/r$, since a significant 
fraction of that energy went into internal energy of the gas by viscous 
dissipation, and little of it got lost by radiation. The gas is therefore only 
marginally bound in the gravitational potential. This suggests that perhaps a 
part of the accreting gas can escape, producing an outflow or wind. In the case 
of the ion supported optically thin ADAFs, this wind would be thermally driven 
by the temperature of the ions, like an `evaporation' from the accretion torus. In 
the case of the radiation supported tori, which exist only at a luminosity near 
the Eddington value, but with much lower temperatures than the ion tori, winds 
driven by radiation pressure could exist.

The possibility of outflows is enhanced by the viscous energy transport through 
the disk. In the case of thin accretion disks (not quite appropriate in the present 
case, but sufficient to demonstrate the effect), the local rate of gravitational energy 
release (erg cm$^{-2}$s$^{-1}$) is $W=\Sigma v_r \p_r(GM/r)$. The local viscous 
dissipation rate is $(9/4)\nu\Sigma\Om^2$. They are related by 
\beq Q_{\rm v}=3[1-({r_{\rm i}\over r})^{1/2}]W, \eeq
where $r_{\rm i}$ is the inner edge of the disk (see `accretion disks' elsewhere 
in this volume). The viscous dissipation rate is less than the gravitational 
energy release for $r<(4/9)r_{\rm i}$, and larger outside this 
radius. Part of the gravitational energy released in the inner disk is transported 
outward by the viscous stresses, so that the  energy deposited in the gas is up 
to three times larger than expected from a local energy balance argument. The 
temperatures in an ADAF would be correspondingly larger. Begelman and 
Blandford (1999) have appealed to this effect to argue that in an ADAF most of 
the accreting mass of a disk might be expelled through a wind, the energy 
needed for this being supplied by the viscous energy transport associated with 
the small amount of mass that actually accretes.

These suggestions are in principle testable, since the arguments are about 
two-dimensional time dependent flows (axisymmetric), which can be studied 
fairly well by numerical simulation. Igumenshchev et al. (1996), and 
Igumenshchev and Abramowicz (1999) present results of such simulations, but 
unfortunately these give a somewhat ambiguous answer to the question. For 
large viscosity ($\alpha\sim 0.3$) no outflow is seen, but for small viscosity 
time dependent flows are seen with outflows in some regions.  Some of these 
flows may be a form of convection and unrelated to systematic outflows.

\end{document}